\documentclass[twocolumn, superscriptaddress, prb, nofootinbib]{revtex4}
\usepackage{graphicx}
\usepackage{amsmath}
\usepackage{multirow}
\usepackage{textcomp}
\usepackage{float}
\usepackage{hyperref}
\usepackage[symbol]{footmisc}
\usepackage{perpage}

\begin{document}
\title{Role of structural H$_2$O in intercalation electrodes: the case of Mg in nano-crystalline Xerogel-V$_2$O$_5$}

\author{Gopalakrishnan Sai Gautam} \email{gautam91@mit.edu}
\affiliation{
Department of Materials Science and Engineering, Massachusetts
Institute of Technology, Cambridge, MA 02139, USA}
\affiliation{
Materials Science Division, Lawrence Berkeley National Laboratory,
Berkeley, CA 94720, USA}

\author{Pieremanuele Canepa}
\affiliation{
Department of Materials Science and Engineering, Massachusetts
Institute of Technology, Cambridge, MA 02139, USA}
\affiliation{
Materials Science Division, Lawrence Berkeley National Laboratory,
Berkeley, CA 94720, USA}

\author{William Davidson Richards}\affiliation{
Department of Materials Science and Engineering, Massachusetts
Institute of Technology, Cambridge, MA 02139, USA}

\author{Rahul Malik}\affiliation{
Department of Materials Science and Engineering, Massachusetts
Institute of Technology, Cambridge, MA 02139, USA}

\author{Gerbrand Ceder} \email{gceder@berkeley.edu, gceder@lbl.gov}
\affiliation{
Materials Science Division, Lawrence Berkeley National Laboratory,
Berkeley, CA 94720, USA}
\affiliation{
Department of Materials Science and Engineering, University of California Berkeley,
CA 94720, USA}



\begin{abstract}
Co-intercalation is a potential approach to influence the voltage and mobility with which cations insert in electrodes for energy storage devices. Combining a robust thermodynamic model with first-principles calculations, we present a detailed investigation revealing the important role of H$_2$O during ion intercalation in nano-materials. We examine the scenario of Mg$^{2+}$ and H$_2$O co-intercalation in nano-crystalline Xerogel-V$_2$O$_5$, a potential cathode material to achieve energy density greater than Li-ion batteries. Water co-intercalation in cathode materials could broadly impact an electrochemical system by influencing its voltages or causing passivation at the anode. The analysis of the stable phases of Mg-Xerogel V$_2$O$_5$ and voltages at different electrolytic conditions reveals a range of concentrations for Mg in the Xerogel and H$_2$O in the electrolyte where there is no thermodynamic driving force for H$_2$O to shuttle with Mg during electrochemical cycling. Also, we demonstrate that H$_2$O shuttling with the Mg$^{2+}$ ions in wet electrolytes yields higher voltages than in dry electrolytes. The thermodynamic framework used to study water and Mg$^{2+}$ co-intercalation in this work opens the door for studying the general phenomenon of solvent co-intercalation observed in other complex solvent-electrode pairs used in the Li- and Na-ion chemical spaces.  
\end{abstract}

\maketitle
\section{Introduction}
\label{sec:intro}
Several cathode materials that have shown appreciable electrochemical performance in Li- and Na-ion batteries are influenced by the presence of H$_2$O in either the cathode structure or the electrolyte. A few examples of these include the MnO$_2$ polymorphs --distorted-spinel Mn$_2$O$_4$,\cite{Li1994} Hollandite,\cite{Rossouw1992} and Birnessite,\cite{Nam2015a,Nam2015} Tavorite-FeSO$_4$F,\cite{Zhang2015,Tripathi2010} Prussian-blue analogues,\cite{Xiao2015a,Lipson2015} 2D Nb/V carbides,\cite{Naguib2013} and Xerogel-V$_2$O$_5$.\cite{Yao1992,Smyrl2001,Wang2006a,Zakharova2007a,Tepavcevic2012} While it is speculated that structural H$_2$O increases the mobility of the intercalating redox-active cation by solvation,\cite{Novak1993,Levi2010b} a key challenge has been to establish whether the structural H$_2$O stays in the electrode or perhaps shuttles with the cation during electrochemical cycling. More generally, the co-intercalation of solvent molecules in layered materials has recently been a focus of great research activity; for example, the thermodynamically prohibited intercalation of Na$^+$ in graphitic anode electrodes is made possible by solvent co-intercalation,\cite{Kim2015d} while the spinel~$\rightarrow$~layered phase transition in MnO$_2$ electrodes is facilitated by water intercalation.\cite{Kim2015e}

Determining how the presence or co-intercalation of water in an electrode influences the intercalation of cations may help to explain contrasting phenomena such as high capacities in a few intercalation systems\cite{Imhof1999,Nam2015,Kim2015c,Sun2015} and rapid capacity fade in a few others,\cite{Rasul2012,Arthur2014,Kim2015b} when water is present. In this study, we investigate the role that H$_2$O plays in the intercalation of Mg$^{2+}$ in nanocrystalline Xerogel-V$_2$O$_5$. Using first-principles calculations, we demonstrate that water co-intercalation with Mg$^{2+}$ is different in wet and dry electrolytes and generally increases the Mg insertion voltage.

While replacing Li$^+$ with a multi-valent ion, such as Mg$^{2+}$ coupled with a Mg metal anode, is viewed as a potential way to achieve higher energy densities than current Li-ion batteries,\cite{Noorden2014,Liu2014a,Rong2015} obtaining cathode materials that can reversibly intercalate Mg$^{2+}$ at high voltage and with substantial capacity remains a pressing challenge.\cite{Shterenberg2014a,Yoo2013,Levi2010b} As a known Li-intercalation host,\cite{Delmas1994} and being one of the few cathode materials that has shown reversible electrochemical Mg$^{2+}$ intercalation,\cite{SaiGautam2015,SaiGautam2015a,Amatucci2001a,Novak1993,Gershinsky2013,Aurbach2000a,Kim2015c,Nam2015} V$_2$O$_5$ is a key component in the design of future multi-valent cathodes. Although orthorhombic-V$_2$O$_5$ possesses multiple polymorphs,\cite{Delmas1994} the nanocrystalline bilayered form of Xerogel-V$_2$O$_5$ is expected to have good Mg mobility owing to electrostatic shielding of the divalent Mg$^{2+}$ by the water contained in the structure.\cite{Novak1993,Levi2010b}

Electrochemical experiments intercalating Mg$^{2+}$ in the Xerogel have reported varying voltages and capacities when employing organic\cite{Imamura2003,Imamura2003a,Tepavcevica2015,Lee2014b} and aqueous\cite{Stojkovic2010,Vujkovic2015} electrolytes, respectively. Imamura \emph{et al.}\cite{Imamura2003,Imamura2003a} showed Mg insertion in Xerogel-V$_2$O$_5$ using acetonitrile (AN) at voltages and capacities higher than that observed with the orthorhombic form \cite{Amatucci2001a,Gershinsky2013} with cyclic performance up to $\sim$~40 cycles at a current density of $\sim$~17~mA/g. Tepavcevic \emph{et al.}\cite{Tepavcevica2015} explored a full-cell arrangement consisting of a Sn anode, Mg(ClO$_4$)$_2$ dissolved in an AN electrolyte and a magnesiated Xerogel cathode and showed reversible Mg intercalation limited by anode capacity. Lee \emph{et al.}\cite{Lee2014b} compared the electrochemical performance of AN and an ethylene carbonate: dimethyl carbonate (EC:DMC) mixture as solvents for Mg insertion and reported improved kinetics with AN than EC:DMC. Stojkovi\'{c} \emph{et al.}\cite{Stojkovic2010} demonstrated reversible Mg intercalation in aqueous electrolytes with a capacity of $\sim$~107~mAh/g at a higher initial voltage (voltage peaks at $\sim$~3.02~V and 2.42~V) compared to the experiments with organic electrolytes.\cite{Imamura2003,Tepavcevica2015} Recently, Vujkovi\'{c} \emph{et al.}\cite{Vujkovic2015} reported high capacity retention ($\sim$~30 cycles) for Mg$^{2+}$ cycling in Xerogel under aqueous electrolytes in comparison to Li$^+$, Na$^+$ or K$^+$. 

So far, there have been no theoretical studies undertaken on the Mg-Xerogel V$_2$O$_5$ system to reveal the role of water co-intercalation under different solvent conditions. In the present work, we describe the Xerogel-V$_2$O$_5$ structure, the phase diagram at 0~K, and voltages as a function of both Mg and H$_2$O content in the structure. We investigate whether the structural H$_2$O in the Xerogel shuttles with the Mg$^{2+}$ ion during cycling at various electrolytic conditions and Mg concentrations in the structure. Finally, we have explored the importance of electrochemical systems with solvent co-intercalation into electrodes, leading to solvent-based voltages that can impact the design of future electrolyte-electrode systems.


\section{Methods}
\label{sec:methods}
To study the effect of H$_2$O on Mg intercalation and understand possible co-intercalation of H$_2$O with the Mg ions, we equilibrate the Mg-Xerogel V$_2$O$_5$ system open to varying amounts of H$_2$O in the electrolyte as governed by the grand-potential, $\Phi = G_{\rm{Mg\text{-}V_{2}O_{5}}} - n_{\rm H_{2}O} . \mu_{\rm H_{2}O}$, with $G_{\rm{Mg\text{-}V_{2} O_{5}}}$, $n_{\rm H_{2}O}$ and $\mu_{\rm H_{2}O}$ the Gibbs energy of the Mg-Xerogel V$_2$O$_5$, the number of moles of water in the Xerogel and the chemical potential of H$_2$O in the electrolyte, respectively. Grand-potential phase diagrams have been used to study open electrochemical systems before.\cite{Ong2008,Piero2015b} While we use Density Functional Theory (DFT, see later)\cite{Kohn1965} calculations to obtain values of $G_{\rm Mg\text{-}V_{2}O_{5}}$ at different Mg concentrations in the Xerogel structure, the procedure used to obtain an accurate reference state for water ($\mu_{\rm H_{2}O}$) is detailed in the Supporting Information (SI).\footnote{$^\#$ Electronic Supporting Information available free of charge online at \url{http://dx.doi.org/10.1021/acs.nanolett.5b05273}}$^{\#}$

DFT calculations are performed with the Vienna Ab Initio Simulation Package,\cite{Kresse1993,Kresse1996} employing the Projector Augmented Wave theory \cite{Kresse1999} with an energy cut-off of 520~eV for describing the wave functions sampled on a well-converged \emph{k}-point (6$\times$2$\times$2) mesh. A Hubbard \emph{U} correction of 3.1~eV is added to remove the spurious self-interaction of the vanadium \emph{d}-electrons.\cite{Anisimov1991,Zhou2004_U,Jain2011} For calculating voltages and phase diagrams at 0~K, the Perdew-Burke-Ernzerhof functional\cite{Perdew1996} in the Generalized Gradient Approximation (GGA) is employed.  Since layered materials such as Xerogel-V$_2$O$_5$ are bound by van der Waals interactions that are not well captured by standard DFT,\cite{Amatucci1996,French2010} the vdW-DF2+\emph{U} functional\cite{Lee2010,Klime2011} is used to compute the layer spacing values ($b-$axis in Figure~\ref{fig:1}). However, preliminary investigations\cite{Carrasco2014,SaiGautam2015,SaiGautam2015a} have shown that GGA+\emph{U} describes the energetics of redox reactions in layered materials better than vdW-DF2+\emph{U}. 

\section{Structure}
\label{sec:Structure}
\begin{figure*}[t]
\includegraphics[scale=0.27]{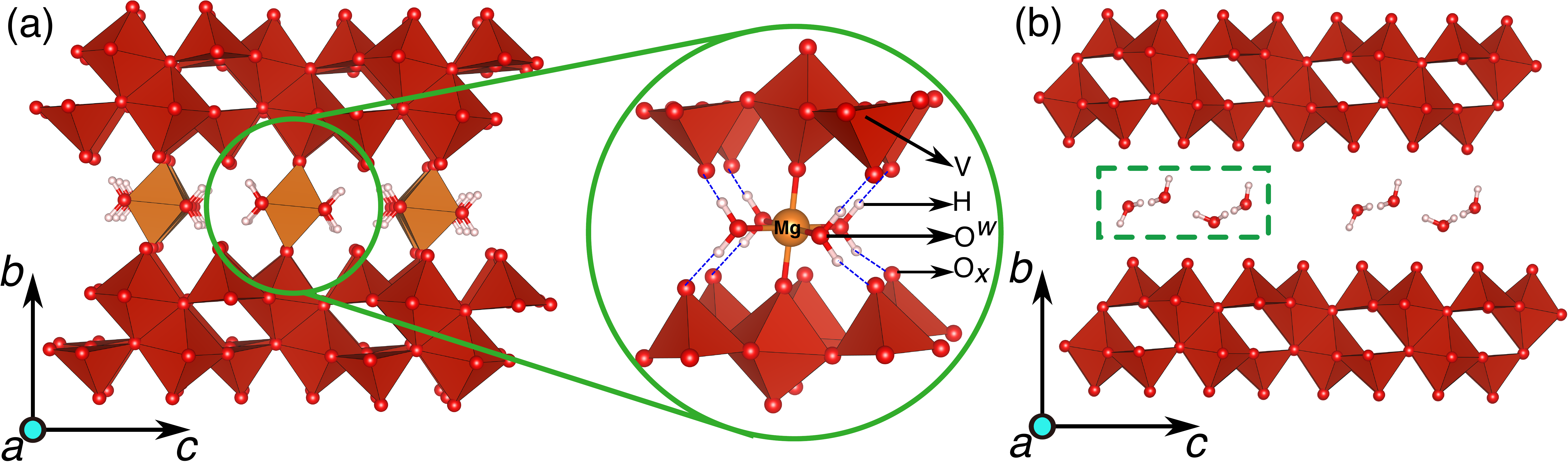}
\caption{
\label{fig:1} 
Structures of the fully magnesiated (x$_{\rm Mg}$~=~0.5) and the fully demagnesiated Xerogel, with 1 H$_2$O per formula unit of V$_2$O$_5$ are displayed in (a) and (b) respectively. The coordination of each Mg by 4 O$^w$ can be observed in the enlarged image in the green circle, with the dashed blue lines indicating hydrogen-bonding between the water molecules and the lattice oxygen. The atomic species in the Xerogel structure are labelled in the enlarged image with O$^w$ and O$_x$ indicating the water and lattice oxygen, respectively.}
\end{figure*}

As the Mg- and H-positions in the Mg-intercalated Xerogel structure are not known experimentally, we have combined relevant experimental information with DFT calculations  to obtain for the first time an atomic-level structural description of this system. While Petkov \emph{et al.}\cite{Petkov2002} resolved the Xerogel-V$_2$O$_5$ structure by employing pair distribution functions from X-ray diffraction measurements, the positions of the intercalant atom were not reported. Oka \emph{et al.}\cite{Oka1999} described the Mg sites in $\sigma$-V$_2$O$_5$, which has a bilayered arrangement but is different from the Xerogel-V$_2$O$_5$ structure. In order to describe not only the Mg (intercalant) positions in the Xerogel structure but also the positions of the water molecules, comprising the oxygen (O$^w$) and the hydrogen atoms, we choose the Ni-intercalated bilayered V$_2$O$_5$ structure as a template (see Figure~\ref{fig:1}).\cite{Oka1997a} As Ni and Mg have similar octahedral coordination preference,\cite{Brown:tr0200} the initial positions of the Mg atoms are obtained from the known Ni-positions in the bilayered structure.\cite{Oka1997a} In this structure, Ni (Mg) is coordinated by 2 oxygen atoms from the VO$_5$ pyramids and 4 O$^w$ atoms as shown in Figure~\ref{fig:1}. The H-positions for the intercalated water in the Xerogel are initialized using a 3-step strategy by placing H-atoms $\sim$~1~\AA{} away from the O$^w$ as explained in the SI and then relaxing these structures with DFT.

Figure~\ref{fig:1}a and b display the fully relaxed structures of the fully magnesiated and demagnesiated Xerogel respectively, containing 1 H$_2$O per formula unit of V$_2$O$_5$ (i.e., $n_{\text{H}_{2}\text{O}}$~=~1) on the $b-c$ plane. Two individual V$_2$O$_5$ layers (red polyhedra in Figure~\ref{fig:1}) are bound by long interlayer V$-$O bonds ($\sim$~2.66~\AA{}) in the fully demagnesiated structure that lead to the formation of a single `bilayer' of V$_2$O$_5$, giving rise to the term ``bilayered'' V$_2$O$_5$. While each V$_2$O$_5$ bilayer is composed of both VO$_5$ square pyramids and VO$_6$ octahedra, the intercalant atoms and the H$_2$O molecules are found in the space between two bilayers. In this work the $a-$, $b-$, and $c-$axes indicate the shortest axis, the inter-bilayer spacing direction, and the longest axis, respectively. 

The orange octahedra in Figure~\ref{fig:1}a correspond to Mg atoms (at the center) coordinated by 6 oxygen atoms. As illustrated in the enlarged version of the Mg coordination environment (green circles, Figure~\ref{fig:1}a), each Mg is bonded to 4 O$^w$ atoms and 2 O atoms of the VO$_x$ polyhedra (referred to as `lattice' oxygen). While H atoms (in white) are bound to O$^w$, the dashed blue lines in Figure~\ref{fig:1}a indicate hydrogen-bonding between the water molecules and the lattice oxygen. The influence of H$_2$O molecules on the electronic structure and density of states in the Mg-Xerogel system is examined in Section~8 of the SI.

On Mg removal, hydrogen-bonding becomes more prominent amongst the H$_2$O molecules than with the lattice oxygen, as deduced by the shorter O--H bonding distances ($\sim$~1.6 -- 1.8~\AA{}) between H and next-nearest O$^w$ atoms compared to hydrogen and lattice oxygen ($\sim$~2.2 -- 2.6~\AA{}), leading to the formation of stable hydrogen-bonded arrangements in the $a-$ and $c-$ directions (dashed green square in Figure~\ref{fig:1}b). The Xerogel structure in our work is limited to 2 fully occupied Mg sites for every 8 vanadium sites, hence the maximum Mg content in the structure is x$_\text{Mg}$~=~0.5 per formula unit of V$_2$O$_5$, hereafter referred to as the ``fully magnesiated" state. Based on our observations in the Ni-based Xerogel structure, we assumed a maximum of 4 H$_2$O molecules for 8 vanadium sites, and $n_{\text{H}_{2}\text{O}}$~=~1 (per V$_2$O$_5$ formula unit) is denoted as the ``fully hydrated" state. 

The inter-bilayer spacing for the fully magnesiated phase (at $n_{\text{H}_{2}\text{O}}$~=~1) using GGA+\emph{U} is $\sim$~10.18~\AA{}, which agrees well with $\sim$~10.22~\AA{} predicted by the vdW-DF2+\emph{U} functional, and is similar to the experimental value of $\sim$~10.36~\AA{} reported for the Ni-intercalated phase.\cite{Oka1997a} The Mg and H$_2$O positions calculated by GGA+\emph{U} and vdW-DF2+\emph{U} are similar, suggesting that the Mg-O electrostatic interactions dominate the geometry of the bilayer once Mg is inserted. For the fully demagnesiated phase (at $n_{\text{H}_{2}\text{O}}$~=~1), the inter-bilayer spacing computed by GGA+\emph{U} ($\sim~12.76~\textrm{\AA}$) differs significantly from the vdW-DF2+\emph{U} value ($\sim$~11.28~\AA{}) and the experimental value of $\sim$~11.52~\AA{}.\cite{Petkov2002} Although GGA+\emph{U} overestimates the layer spacing for the fully demagnesiated Xerogel structure (at $n_{\text{H}_{2}\text{O}}$~=~1, as in Figure~\ref{fig:1}b), the hydrogen-bonded arrangement of H$_2$O molecules is similar to that found with the vdW-DF2+\emph{U} functional.

\section{Equilibration of the water content}
\label{sec:phase-diagrams}
Obtaining the equilibrium water content in the Xerogel requires one to know the free energy of the cathode as a function of the water content in the electrolyte, after which a minimization of the grand potential at the $\mu_{\rm H_{2}O}$ of the electrolyte gives the equilibrium amount of H$_2$O in the cathode. We calculated the free energies of various Xerogel structures, enumerated in supercell volumes twice that of the conventional cell. We assessed the stability of the enumerated structures at $\rm{x_{Mg}} = 0, 0.25$ and 0.5, containing various amounts of co-intercalated H$_2$O ($n_{\rm{H_{2}O}} = 0, 0.5$ and 1), and for several water-concentration regimes in the electrolyte (see Figure~S1 in the SI). 

\begin{figure}[H]
\includegraphics[width=\columnwidth]{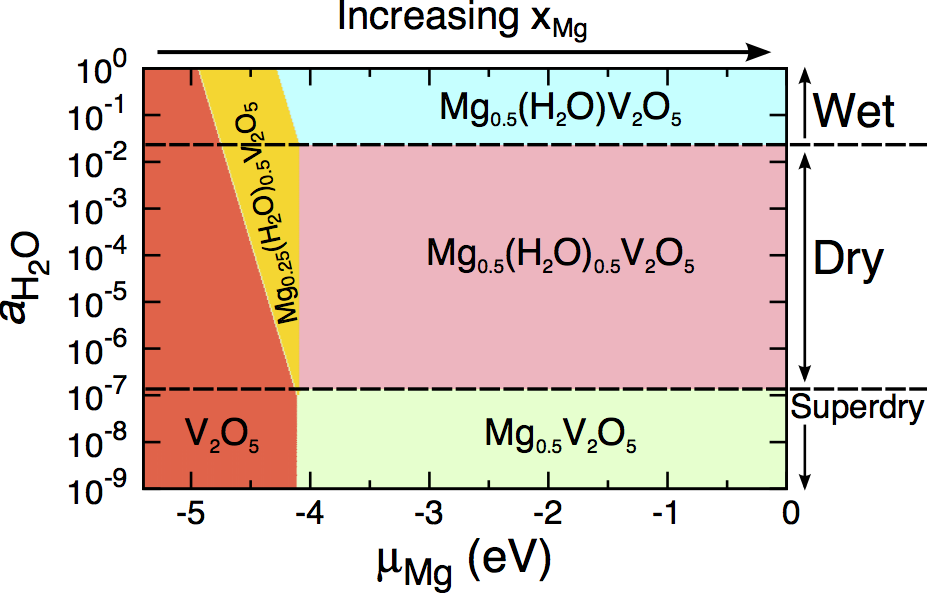}
\caption{
\label{fig:2} 
The Grand-potential phase diagram at 0~K of Mg-Xerogel V$_2$O$_5$ as a function of various electrolytic conditions and Mg chemical potentials is shown. Each colored region represents a single phase with the indicated Mg and water content. The dashed lines display different electrolytic regimes, with $\mu_{\rm Mg} =$~0 corresponding to full magnesiation.} 
\end{figure}

The stable Mg-Xerogel V$_2$O$_5$ phases, obtained by minimizing the grand-potential at 0~K, are plotted in Figure~\ref{fig:2} as a function of $a_{\rm H_{2}O}$ and the Mg chemical potential ($\mu_{\rm Mg}$) with pictorial descriptions provided in Figure~S2 in the SI. A high Mg chemical potential, such as $\mu_{\rm Mg}=0$ (see Section~6, SI), refers to a highly magnesiated Xerogel configuration (x$_{\rm Mg} =$~0.5), while decreasing the chemical potential represents a more oxidizing environment that leads to demagnesiation (x$_{\rm Mg} \sim 0$). To explore the effect of changing electrolytic conditions on the electrochemical properties of Xerogel-V$_2$O$_5$, we consider three different regimes (separated by dashed lines in Figure~\ref{fig:2}): $i$) $wet$ or aqueous electrolyte, where the water activity, $a_{\rm{H_{2}O}}$, is set to $\sim$~1, $ii$) $dry$ with 10$^{-2} < a_{\rm{H_{2}O}} <$~10$^{-6}$, and $iii$) $superdry$ with $a_{\rm{H_{2}O}} <$~10$^{-7}$. An activity $a_{\rm{H_{2}O}} =$~10$^{-4}$ would correspond to $\approx$~10~ppm by weight of water under the ideal solution approximation in solvents such as glymes\cite{Ramkumar1985,Tang2014}. Each colored region in Figure~\ref{fig:2} corresponds to a single stable phase, whose composition is indicated with a Mg$_{\rm x}$(H$_2$O)$_n$V$_2$O$_5$ notation. While the lines separating the single phase regions indicate the co-existence of two phases, the triple points correspond to a three phase co-existence. 

For a wet electrolyte ($a_{\text{H}_{2}\text{O}} \sim$~1), the ground state structures as a function of $\mu_{\rm Mg}$ consist of the fully magnesiated $-$ fully hydrated structure (x$_\text{Mg}$~=~0.5, $n_{\text{H}_{2}\text{O}}=1$ per V$_2$O$_5$, blue region in Figure~\ref{fig:2}), the `half' magnesiated $-$ half hydrated structure (x$_\text{Mg}$~=~0.25, $n_{\text{H}_{2}\text{O}}=0.5$, yellow region), and the fully demagnesiated $-$ dehydrated structure (x$_\text{Mg}$~=~0, $n_{\text{H}_{2}\text{O}}$~$=$~0, red region). Hence, under these electrolyte conditions, each Mg$^{2+}$ intercalates with two H$_2$O molecules and a decrease in Mg content also corresponds to a decrease of water intercalated. Thus, when an aqueous electrolyte is used, there is a thermodynamic driving force for the water content to change with the Mg content.

When Mg$^{2+}$ intercalation occurs from a dry electrolyte (10$^{-2} < a_{\text{H}_{2}\text{O}} <$~10$^{-6}$), the ground state phases are: x$_\text{Mg}$~=~0.5, $n_{\text{H}_{2}\text{O}}=~0.5$ (fully magnesiated $-$ half hydrated, pink region in Figure~\ref{fig:2}), x$_\text{Mg}$~=~0.25, $n_{\text{H}_{2}\text{O}}$~=~0.5 (half magnesiated $-$ half hydrated, yellow), and x$_\text{Mg}$~=0, $n_{\text{H}_{2}\text{O}}$~=~0 (fully demagnesiated $-$ dehydrated, red). The results demonstrate that in a dry electrolyte, H$_2$O co-intercalates with Mg for x$_{\rm Mg} <$~0.25, whereas the water content remains unchanged as more Mg is inserted.

For a superdry electrolyte ($a_{\text{H}_{2}\text{O}} <$~10$^{-7}$) the stable phases consist of fully dehydrated structures, both at x$_\text{Mg}$~=~0.5 (fully magnesiated, green region in Figure~\ref{fig:2}) and x$_\text{Mg}$~=~0 (fully demagnesiated, red). The absence of ground state configurations at intermediate Mg compositions (Figure~S1c in the SI) in a superdry electrolyte indicates a phase separating behavior into Mg-rich and Mg-poor domains. Since the superdry ground states are fully dehydrated, there is a high driving force for all the water in the Xerogel to leave the structure. Interestingly, the activity of H$_2$O in the electrolyte not only influences the level of co-intercalation but also controls the nature of the Mg intercalation. Without water Mg-intercalation occurs as a 2-phase reaction between x$_{\rm Mg} =$~0 and x$_{\rm Mg} =$~0.5, whereas water in the electrolyte stabilizes intermediate Mg states.

The ground state structure of V$_2$O$_5$, across  the range of $\mu_{\rm Mg}$ and $a_{\rm H_{2}O}$ considered is the orthorhombic $\alpha$-V$_2$O$_5$, \cite{Delmas1994,SaiGautam2015} which is consistent with experimental evidence of an irreversible transformation of the Xerogel to $\alpha$-V$_2$O$_5$ at high temperatures,\cite{Legendre1983} suggesting the metastable nature of the Xerogel. In fact, the $\alpha$ polymorph is lower in energy at x$_\text{Mg}$~=~0 and 0.5 compared to the dehydrated Xerogel phases (red and green regions in Figure~\ref{fig:2}) by $\sim$~360~meV/f.u. and $\sim$~200~meV/f.u., respectively.

Combining the results of Figure~\ref{fig:2}, we find that under wet conditions Mg$^{2+}$ ions shuttle along with H$_2$O molecules across Mg concentrations, whereas under dry conditions H$_2$O co-intercalation only occurs between $0 \leq {\rm x_{Mg}} \leq 0.25$. Hence, water will not shuttle with Mg under dry conditions and high Mg concentrations ($0.25 \leq \text{x}_\text{Mg} \leq 0.5$) in the Xerogel.  In a superdry electrolyte, there is no H$_2$O within the Xerogel structure. Although we have discussed the general phenomenon of Mg-H$_2$O co-intercalation\cite{Nam2015,Novak1993} for the case of Xerogel-V$_2$O$_5$, similar models are readily applicable to study solvent co-intercalation in other layered electrode materials.\cite{Kim2015d}

\section{Effect of water on the Mg insertion voltage}
\label{sec:Voltages}
\begin{figure}[h]
\includegraphics[width=\columnwidth]{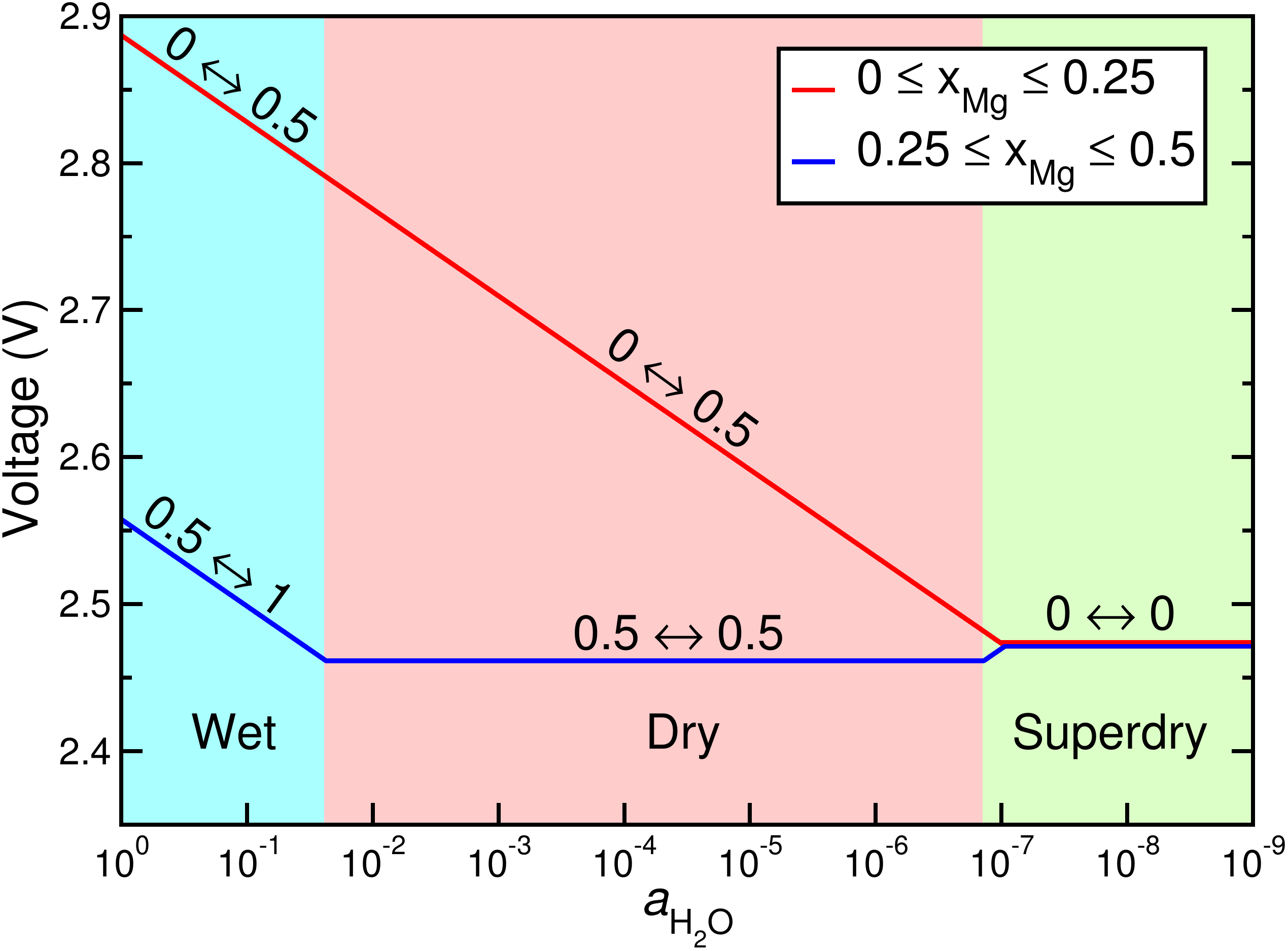}
\caption{
\label{fig:3}
Average Mg insertion voltage for low (red line) and high (blue) Mg concentrations as a function of the electrolyte water content ($a_{\text{H}_{2}\text{O}})$. Equations on the curves indicate the change in H$_2$O content in the Xerogel as Mg is inserted in each electrolytic regime.}
\end{figure}

In regimes where H$_2$O shuttles with the Mg, the activity of water affects the cell voltage, as illustrated by the average voltage curves computed for Mg insertion into Xerogel-V$_2$O$_5$ in Figure~\ref{fig:3}. The voltages as a function of $a_{\rm{H_{2}O}}$, under the wet (cyan background), dry (pink), and superdry (green) regimes, are obtained from the phase diagram of Figure~\ref{fig:2}, using the procedure detailed in the SI.\cite{Aydinol1997} The red and blue lines indicate the voltages for Mg insertion between concentration ranges of $0 \leq \text{x}_\text{Mg} \leq 0.25$ and $0.25 \leq \text{x}_\text{Mg} \leq 0.5$ respectively. Thus, at a given $a_{\text{H}_{2}\text{O}}$, the values on the red and blue curves indicate the average voltage that will be observed between $0 \leq \text{x}_\text{Mg} \leq 0.25$ and $0.25 \leq \text{x}_\text{Mg} \leq 0.5$. The equations on the voltage curves indicate  changes in the structural H$_2$O content of the Xerogel, as Mg is inserted. For example, ``$0.5 \leftrightarrow 1$" on the blue line for $a_{\text{H}_{2}\text{O}} \sim$~10$^{-1}$ (wet electrolyte) indicates a variation in $n_{\text{H}_{2}\text{O}}$ from 0.5 to 1 as x$_\text{Mg}$ increases from 0.25 to 0.5. The slope changes of the voltage curves, particularly the ones at high Mg concentration (blue line), indicate the critical water content in the electrolyte at which the H$_2$O co-intercalation behavior changes. The merging of the red and blue curves in the superdry region in Figure~\ref{fig:3} reflects that only a single voltage plateau for $0 \leq \text{x}_\text{Mg} \leq 0.5$ is found. The interpretation of the kink on the blue voltage curve observed in the superdry region is given in the SI.

Although Mg intercalation experiments in Xerogel-V$_2$O$_5$ are normally performed on structures with higher H$_2$O and Mg content than considered in our structural model,\cite{Legendre1983,Imamura2003,Imamura2003a, Tepavcevica2015} the calculated voltage curves in Figure~\ref{fig:3} qualitatively agree with the experimental voltage features for Mg insertion in wet,\cite{Stojkovic2010} and dry electrolytes.\cite{Imamura2003} The calculated voltage for the superdry electrolyte ($\sim$~2.47~V, $a_{\text{H}_{2}\text{O}} \sim 10^{-8}$), where the H$_2$O exits the Xerogel during Mg cycling, is higher but comparable to $\alpha$-V$_2$O$_5$ at low Mg concentrations ($\sim$~2.44~V). \cite{SaiGautam2015} Importantly, the increase in voltages with increase in $a_{\rm{H_{2}O}}$, as predicted by theory (Figure~\ref{fig:3}), is in good agreement with experimental observations of higher initial voltages in aqueous (voltage peak at $\sim$~3.02~V) compared to dry (peak at $\sim$~2.88~V) electrolytes and $\alpha$-Mg$_\text{x}$V$_2$O$_5$ ($\sim$~2.35~V, no water).\cite{Gershinsky2013,Stojkovic2010, Imamura2003}
  

\section{Discussion}
\label{sec:discussion}
In this work, we have used first-principles methods based on DFT to investigate Mg intercalation into Xerogel-V$_2$O$_5$. Specifically, we have clarified the structure of the Xerogel, evaluated the phase diagrams for Mg intercalation under different electrolytic conditions (wet, dry and superdry), and calculated the average voltages for each case. The data presented in this work not only sheds light on existing experiments in the Mg-Xerogel system, with possible Mg-H$_2$O co-intercalation, but also provides a working model for studying solvent co-intercalation properties in layered materials for batteries and other applications.

\begin{figure*}[!th]
\includegraphics[scale=0.8]{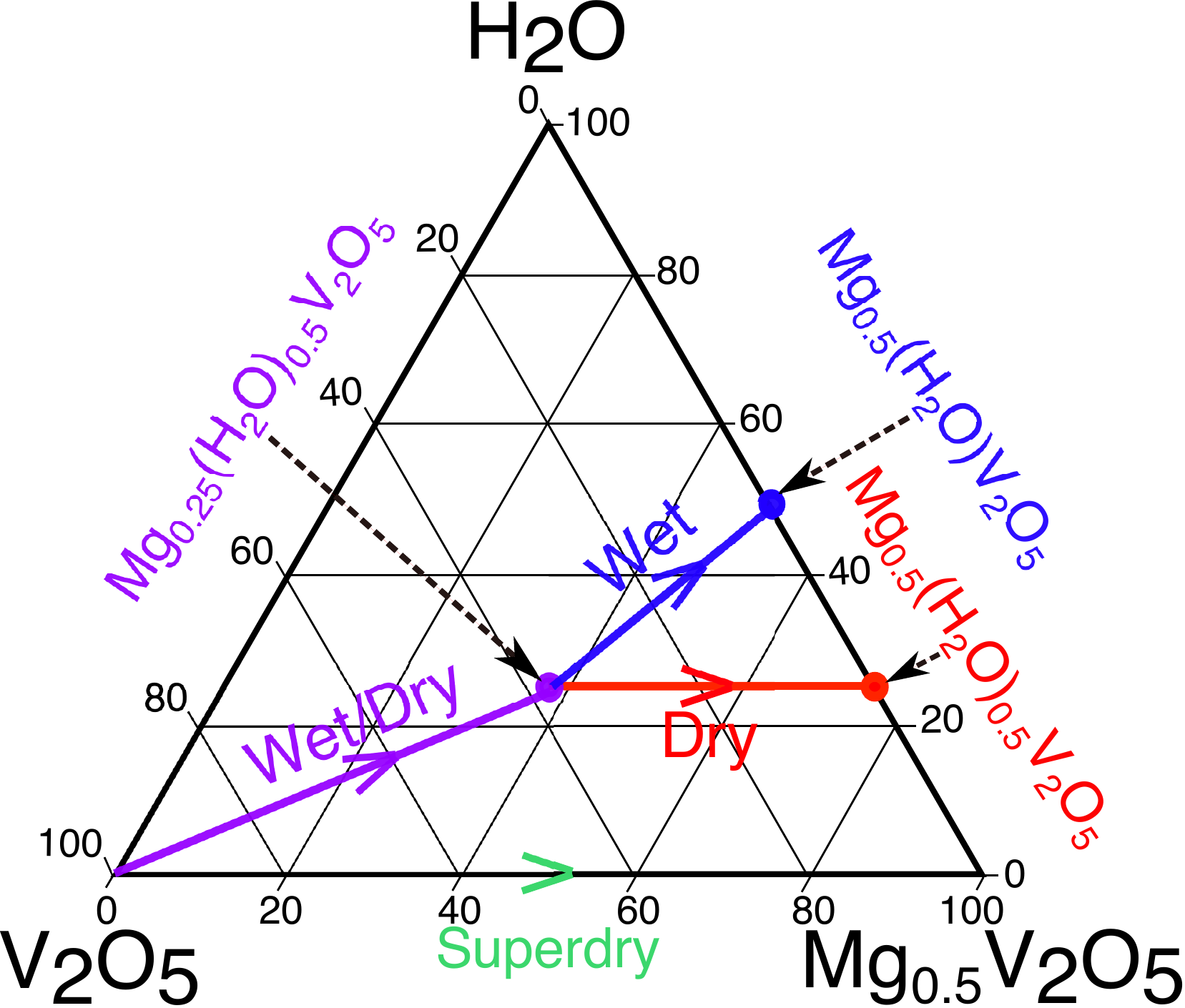}
\caption{
\label{fig:4}
Ternary phase diagram of the Mg-(Xerogel)V$_2$O$_5$-H$_2$O system, which summarizes the possible equilibrium phases under different electrolyte conditions. The ``Wet/Dry" trajectory indicates that the equilibrium states are similar for both wet and dry electrolytes in that Mg concentration range. The green arrow shows the stable phases in a superdry electrolyte.}
\end{figure*}

Figure~\ref{fig:4} displays a phase diagram of the Xerogel V$_2$O$_5$~$-$~Magnesiated Xerogel V$_2$O$_5$~$-$~H$_2$O ternary system, summarizing the results of Figure~\ref{fig:2}. The base of the triangle (Figure~\ref{fig:4}) corresponds to Mg intercalation in the Xerogel-V$_2$O$_5$ structure in the absence of H$_2$O, or the superdry electrolyte, as indicated by the green arrow. The colored solid lines on the phase diagram represent the trajectories of stable phases that will form upon magnesiation of the Xerogel-V$_2$O$_5$ structure under different electrolyte conditions. While the solid blue and red lines correspond to the wet and the dry electrolytes respectively, the purple line indicates the commonality of the stable phases between wet and dry electrolytes at low Mg concentrations. The blue and red circles are the stable states at full magnesiation in a wet and dry electrolyte, respectively. The purple circle indicates the half magnesiated $-$ half hydrated ground state common to both the wet and dry electrolytes. 

While initial Mg intercalation up to x$_{\rm Mg} =$~0.25 pulls H$_2$O into the structure for both wet and dry electrolytes, further co-intercalation of water with Mg depends more sensitively on the water content of the electrolyte. Interestingly, the presence of water in the electrolyte changes the phase behavior of the Mg-Xerogel system from that of a two phase reaction at a single voltage (superdry) to one with a capacity over a range of voltages (wet and dry).

In conventional secondary batteries, where the solvent or electrolyte do not co-intercalate with the redox-active cation, the voltage depends on the chemical potential difference of the cation species between the cathode and the anode.\cite{Aydinol1997,Ceder1999} However, our study suggests that the measured voltages are subjected to change if the co-intercalation of the solvent/electrolyte with the redox ion occurs, leading to a co-dependence on the solvent/electrolyte chemical potential. As illustrated by Figure~\ref{fig:3}, the Mg insertion voltage in the Xerogel is calculated to be $\sim$~150~mV higher in a wet electrolyte than in a dry electrolyte ($a_{\text{H}_{2}\text{O}} \sim$~10$^{-4}$), in good agreement with reported higher voltages in aqueous compared to organic (dry) electrolytes.\cite{Imamura2003,Tepavcevica2015,Stojkovic2010,Vujkovic2015} Electrolyte/solvent-dependent voltages give rise to important design consequences in a battery system, since the voltage generated can be calibrated based on both the solvent polarity (polar/apolar) and the quantity (wet/dry) of the intercalating solvent species. Further analysis on the variability of voltages based on solvents is relevant not only in the design of improved electrolytes but also in selecting possible electrolyte-additive combinations that can ultimately improve the energy density of an electrochemical system.

H$_2$O co-intercalation in Xerogel-V$_2$O$_5$ has three important technological consequences: $i$) higher Mg insertion voltages, $ii$) change in phase behavior from a two phase regime (superdry) to one with intermediate stable Mg concentrations (wet, dry) and $iii$) higher kinetic rate of Mg insertion originating from the electrostatic shielding effect of the coordinating water molecules in the cathode.\cite{Levi2010b,Novak1993} Nevertheless, in the case of Mg-ion batteries, where the Mg metal anode is crucial to achieve energy densities higher than current Li-ion technology,\cite{Noorden2014} the presence of H$_2$O in the electrolyte or coordinated with the Mg$^{2+}$ ions could cause passivation at the Mg anode.\cite{Muldoon2012,Gofer2003a,Shterenberg2014a} While there exist solvents that successfully solvate Mg$^{2+}$ and do not cause passivation of the Mg metal (e.g., ethers like tetrahydrofuran and glymes\cite{Yoo2013}), it is crucial to understand their fate as a co-intercalant together with the Mg in the bilayered-V$_2$O$_5$ structure, and their impact on the Mg insertion voltage and mobility. More generally, investigations of solvent co-intercalation properties in other layered materials will be useful and important in designing the next generation of rechargeable Li, Na and multi-valent batteries.

\section{Conclusion}
\label{sec:conclusion}
In this work, we have integrated experimental information with first-principles computations to resolve the nano-crystalline Mg-Xerogel V$_2$O$_5$ structure and observed Mg being coordinated by 2 lattice oxygen and 4 oxygen from co-intercalated H$_2$O. Using grand-potential phase diagrams, we found that water co-intercalation with Mg$^{2+}$ depends on the water activity in the electrolyte, ranging from full co-intercalation in wet to none in superdry conditions. Also, we have established the significant impact of water (or solvent) co-intercalation on the voltages and voltage profiles obtained.
\\

\begin{acknowledgements}
The current work is fully supported by the Joint Center for Energy Storage Research (JCESR), an Energy Innovation Hub funded by the U.S. Department of Energy, Office of Science and Basic Energy Sciences. This study was supported by Subcontract 3F-31144. The authors thank the National Energy Research Scientific Computing Center (NERSC) for providing computing resources. The authors declare no competing financial interests.
\end{acknowledgements}

\bibliography{library}

\end{document}